\documentclass[12pt]{article}
\usepackage{lmodern}
\usepackage[T1]{fontenc}
\usepackage[utf8]{inputenc}
\usepackage{geometry}
\usepackage{xcolor}
\usepackage{url}
\usepackage{amsmath}
\usepackage{amssymb}
\usepackage{graphicx}
\usepackage{setspace}
\PassOptionsToPackage{version=3}{mhchem}
\usepackage{mhchem}
\usepackage[unicode=true,
 bookmarks=true,bookmarksnumbered=false,bookmarksopen=false,
 breaklinks=true,pdfborder={0 0 0},pdfborderstyle={},backref=false,colorlinks=true]
 {hyperref}
\hypersetup{pdftitle={Comparison of methods},
 citecolor =blue, linkcolor = blue,  urlcolor  = blue}

\makeatletter

\providecommand{\tabularnewline}{\\}

\usepackage{datetime}
\usepackage{refstyle}
\usepackage{url}
\usepackage[title,toc,page,header]{appendix} 
\usepackage[nosort,super]{cite}

\usepackage{todonotes}

\makeatother

\begin{document}

\title{Using the Gini coefficient to characterize the shape \\
of computational chemistry error distributions}

\author{Pascal PERNOT \\
Institut de Chimie Physique, UMR8000 CNRS, \\
Université Paris-Saclay, 91405, Orsay, France\\
Email: pascal.pernot@universite-paris-saclay.fr\\
\&~\\
Andreas SAVIN \\
Laboratoire de Chimie Théorique, \\
CNRS and UPMC Université Paris 06,\\
Sorbonne Universités, F-75252 Paris, France\\
Email: andreas.savin@lct.jussieu.fr}

\maketitle
\pagebreak{}
\begin{abstract}
The distribution of errors is a central object in the assesment and
benchmarking of computational chemistry methods. The popular and often
blind use of the mean unsigned error as a benchmarking statistic leads
to ignore distributions features that impact the reliability of the
tested methods. We explore how the Gini coefficient offers a global
representation of the errors distribution, but, except for extreme
values, does not enable an unambiguous diagnostic. We propose to relieve
the ambiguity by applying the Gini coefficient to mode-centered error
distributions. This version can usefully complement benchmarking statistics
and alert on error sets with potentially problematic shapes.
\end{abstract}
\pagebreak{}

\section{Introduction}

The reliability of a computational chemistry method is conditioned
by the distribution of its prediction errors. Distributions with heavy
tails elicit a risk of large prediction errors. As a benchmarking
statistic, the popular mean unsigned error (MUE) bears no information
on such a risk \cite{Pernot2015,Pernot2018,Pernot2020,Pernot2020a}.
We have recently reported a case where two unbiased error distributions
with identical values of the MUE present widely different risks of
large errors because of heavy tails in one of them \cite{Pernot2020b,Pernot2020a}.
It would therefore be very useful to complement the MUE with a statistic
indicating or quantifying the risk of large errors.

We recently proposed alternative statistics such as $Q_{95}$ \cite{Pernot2018},
$P_{\eta}$, \cite{Pernot2018} and systematic improvement probability
(SIP) \cite{Pernot2020}. In terms of risk, these statistics offer
the following interpretations: 
\begin{itemize}
\item There is a 5\,\% risk for absolute errors to exceed $Q_{95}$.
\item There is a probability $P_{\eta}$ that absolute errors are larger
than a chosen threshold $\eta$. $P_{\eta}$ provides a direct quantification
of the risk of large errors, but $\eta$ has to be defined \emph{a
priori} and might be user-dependent, which complicates its reporting
in benchmarking studies. 
\item For two methods $M_{1}$ and $M_{2}$, the SIP provides the system-wise
probability that the absolute errors of $M_{1}$ are smaller than
the absolute errors of $M_{2}$, informing on the risk incurred by
switching between two methods. Interestingly, the SIP analysis provides
a decomposition of the MUE difference between two methods in terms
of gain and loss probabilities \cite{Pernot2020}. 
\end{itemize}
The $Q_{95}$ and $P_{\eta}$ partly answer the question, but they
are point estimates on the cumulated density function of the absolute
errors, and a statistic accounting for the whole distribution might
be of interest. Besides, it is well established in econometrics that
measures of dispersions such as the variance perform poorly at risk
estimation and that higher moments of the distributions have to be
considered \cite{Bonato2011}. This would lead us to such measures
as skewness and kurtosis, but none of these alone would be able to
cover all the scenarii. The risk of large errors through heavy tails
of the errors distribution might be associated with large skewness
or large kurtosis or a combination of them.

The Lorenz curve \cite{Lorenz1905} is widely used in econometrics
to represent the distribution of wealth in human populations. Its
summary statistics, notably the Gini coefficient (noted $G$) \cite{Gini1912,Damgaard2000,Eliazar2010},
are used to evaluate the degree of inequality within these populations.
The Gini coefficient is also used, for instance, in ecology, to estimate
the inequality of properties within plant populations \cite{Bendel1989,Damgaard2000},
in astronomy, to characterize the morphology of galaxies \cite{Florian2016},
or in information theory, to characterize the sparsity of datasets
\cite{Hurley2009}. 

The Lorenz curve has many mathematical representations, the most interesting
one, for us, being its formulation as an integral of the quantile
function, a direct link with our study of probabilistic metrics \cite{Pernot2018,Pernot2020,Pernot2020a}.
More precisely, we explore here the interest of the Gini coefficient
as a complement to the MUE in benchmarking studies. 

We introduce the statistical tools and their implementation in Section\,\ref{sec:Statistical-methods},
and apply them to a series of datasets to illustrate the interest
and limitations of the Gini coefficient in Section\,\ref{sec:Applications}.
An adaptation of the Gini coefficient is proposed to relieve its main
drawbacks when applied to error datasets.

\section{Statistical methods\label{sec:Statistical-methods}}

\subsection{Definitions}

Let us consider a distribution of errors $e$ with probability density
function (PDF) $f(e)$. The absolute errors $\varepsilon=|e|$ have
a \emph{folded} PDF $f_{F}(\varepsilon)$. To avoid ambiguity, statistics
based on absolute errors are indexed by $F$.

\subsubsection{CDF and quantile function}

The cumulative distribution function (CDF) of the absolute errors
is noted 
\begin{equation}
C_{F}(z)=\int_{0}^{z}f_{F}(\varepsilon)\thinspace d\varepsilon
\end{equation}
from which the quantile function is the inverse
\begin{equation}
q_{F}(p)=C_{F}^{-1}(p)
\end{equation}
\begin{figure}[!t]
\noindent \begin{centering}
\includegraphics[width=0.95\textwidth]{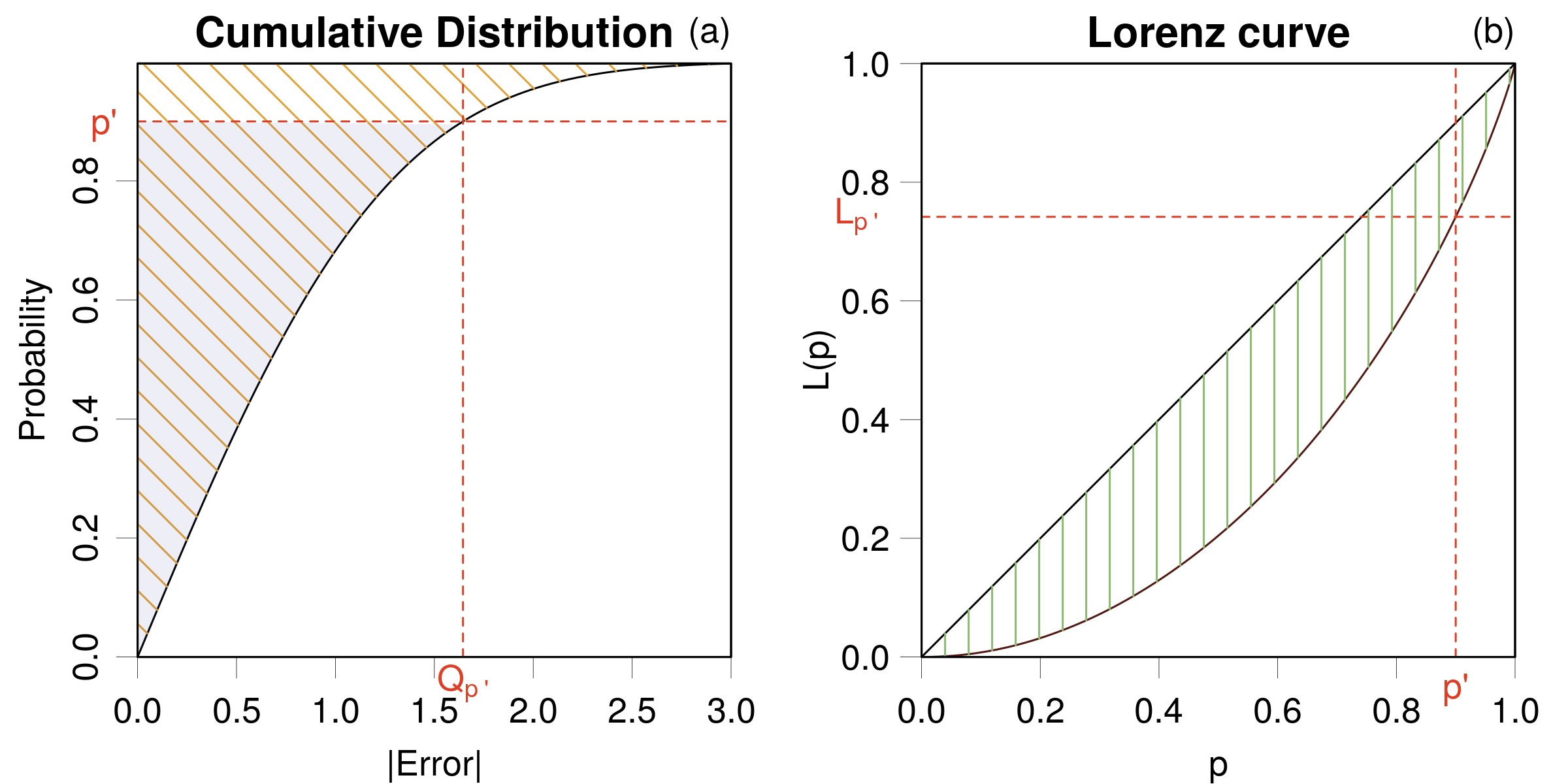}
\par\end{centering}
\caption{\label{fig:principle}Schematic cumulative density function (a) and
Lorenz curve (b) for a folded standard normal density function of
absolute errors. The area above the CDF (slanted) is the mean unsigned
error (MUE). For a given probability $p'$, the ratio of the shaded
area to the total slanted area gives the value of the Lorenz curve
$L_{p'}=L(p')$. $Q_{p'}$ is the quantile for probability $p'$.
The area between the Lorenz curve and the identity axis (vertical
bars) is half the Gini coefficient.}
\end{figure}

\subsubsection{Mean unsigned error}

The mean unsigned error (MUE), is defined as
\begin{align}
\mu_{F} & =\int_{0}^{\infty}\varepsilon f_{F}(\varepsilon)\thinspace d\varepsilon
\end{align}
Using the change of variable $\varepsilon=C_{F}^{-1}(p)$, $p=C_{F}(\varepsilon)$
and $dp=f_{F}(\varepsilon)\thinspace d\varepsilon$, the MUE also
can be shown to be the integral of the quantile function
\begin{align}
\mu_{F} & =\int_{0}^{1}C_{F}^{-1}(p)\thinspace dp\\
 & =\int_{0}^{1}q_{F}(p)\thinspace dp
\end{align}

\subsubsection{The Lorenz curve}

The Lorenz curve gives the percentage of cumulated absolute errors
due to the $100\times p$\,\% smallest values, or equivalently, the
portion of the MUE due to the $100\times p$\,\% smallest absolute
errors: 

\begin{equation}
L_{F}(p)=\frac{1}{\mu_{F}}\int_{0}^{p}q_{F}(t)\thinspace dt
\end{equation}
As shown in Fig.\,\ref{fig:principle}(a), it is the ratio between
the slanted shaded area and the total slanted area. Its value for
$p'$ is reported on the corresponding Lorenz curve graph (Fig.\,\ref{fig:principle}(b)).

The Lorenz curve provides a \emph{scale-invariant} representation
of the CDF $C_{F}(z)$ \cite{Kleiber2005}, with the following properties:
$L_{F}(p)$ is concave, increasing with $p$, such as $0\le L_{F}(p)\le p\le1$,
$L_{F}(0)=0$ and $L_{F}(1)=1$. $L_{F}(p)$ lies one the identity
line ($L_{F}(p)=p$) when all the errors are equal, \emph{i.e.} $f_{F}(\varepsilon)=\delta(\varepsilon-c)$.
Note that this case corresponds to a discontinuous CDF, with a jump
at $\varepsilon=c$. The deviation of an error distribution from this
extreme case can be measured by the Gini coefficient.

\subsubsection{The Gini coefficient\label{subsec:The-Gini-coefficient}}

It is related to the area between $L_{F}(p)$ and the identity line
(Fig.\,\ref{fig:principle}(b))
\begin{equation}
G_{F}=2\int_{0}^{1}\left\{ p-L(p)\right\} \thinspace dp
\end{equation}
where the factor two is used to scale $G_{F}$ between 0 and 1. The
smaller $G_{F}$, the closer the Lorenz curve to the identity line.
The Gini coefficient, usually noted $G$, is generally used for distributions
with positive support. Our notation $G_{F}$ is a reminder that we
are working here with distributions of absolute errors $f_{F}(\varepsilon)$.

For sets of \emph{absolute} errors with a normal distribution $N(\varepsilon;\mu_{F},\sigma_{F})$,
$G_{F}$ is proportional to the coefficient of variation $c_{v}=\sigma_{F}/\mu_{F}$
\cite{Bendel1989}, where $\sigma_{F}$ the standard deviation of
the absolute errors 
\begin{equation}
G_{F}\sim c_{v}/\sqrt{\pi}\label{eq:GcsCv}
\end{equation}
Note that this relationship does hold only when all errors are of
the same sign ($\mu_{F}\gg\sigma_{F}$), therefore with small $c_{v}$
values.

Two typical values of $G_{F}$ will be useful in the following: 
\begin{itemize}
\item for any zero-centered normal distribution of errors $N(0,\sigma)$,
the distribution of absolute errors is the half-normal distribution,
with value $G_{FN}=\sqrt{2}-1\simeq0.41$ \cite{Dixon1987}; 
\item for any zero-centered uniform distribution, $U(-a,a)$, or any uniform
distribution with a bound at zero, $U(-a,0)$ or $U(0,a)$, folding
produces a uniform distribution with the minimal bound at zero, with
value $G_{FU}=1/3$ \cite{Dixon1987}. 
\end{itemize}

\subsubsection{Skewness and kurtosis}

Skewness measures the asymmetry of a distribution, while kurtosis
is used as a measure either of its ``peakedness'' or ``tailedness''
\cite{Ruppert1987} The moments-based formulae for skewness and kurtosis
are not robust to outliers, and more robust quantile-based formulae
have been proposed by several authors \cite{Groeneveld1984,Ruppert1987,Bonato2011,Suaray2015}.

For the skewness, we use a measure using the difference between the
mean and median
\begin{equation}
\beta_{GM}=\frac{\mu-q(0.5)}{<|e-q(0.5)|>}\label{eq:skewgm}
\end{equation}
where the brackets indicate the mean value, $q(0.5)$ is the median
of \emph{signed} error, $e$, and the $GM$ subscript refers to the
authors of this definition, Goeneveld and Meeden \cite{Groeneveld1984}.
$\beta_{GM}$ takes its values between -1 and 1, and is 0 for symmetric
distributions. 

For kurtosis, an estimate based on quantiles is used \cite{Bonato2011}
(originating from a similar form proposed by Crown and Siddiqui \cite{Crow1967},
hence the $CS$ subscript)
\begin{equation}
\kappa_{CS}=\frac{q(0.975)-q(0.025)}{q(0.75)-q(0.25)}-2.91\label{eq:kurtcs}
\end{equation}
where $q(.)$ is the quantile function for \emph{signed} errors. The
correction factor for the normal distribution (2.91) makes that $\kappa_{CS}$
measures an \emph{excess} kurtosis. Datasets with $\kappa_{CS}>0$
have heavier tails than a normal distribution, and the opposite for
negative values.

Specific notations will be introduced below when these statistics
are applied to sets of absolute errors. 

\subsection{Estimation}

We consider in this section the application of the previous statistics
to finite error samples, and the corresponding formulae. Let us consider
a set of errors ($E=\left\{ e_{i}\right\} _{i=1}^{N}$), derived from
a set of $N$ calculated values ($C=\left\{ c_{i}\right\} _{i=1}^{N}$)
and reference data ($R=\left\{ r_{i}\right\} _{i=1}^{N}$), by $e_{i}=r_{i}-c_{i}$.
The absolute errors are noted $\varXi=\left\{ \varepsilon_{i}=|e_{i}|\right\} _{i=1}^{N}$. 

\paragraph{MSE, MUE and mode.}

The mean signed error (MSE) is estimated as $\mu=\frac{\text{1}}{N}\sum_{i=1}^{N}e_{i}$,
and the mean unsigned error (MUE) as $\mu_{F}=\frac{\text{1}}{N}\sum_{i=1}^{N}\varepsilon_{i}$. 

As one is not dealing with necessarily symmetric distributions, the
mode is an interesting location statistic, notably in correspondence
to the tails of a distribution. The mode locates the part of the population
with the highest density, which is expected to bring a large contribution
to $\mu_{F}$, and therefore influence the Lorenz curve and Gini coefficient.
As one cannot assume the unimodality of the underlying distributions,
one will consider the main mode. A non-parametric robust method, Bickel's
half-range mode (HRM) estimator \cite{Bickel2002,Hedges2003}, is
used to estimate the location of the error samples main mode. This
methods proceeds by iterative bipartition of modal intervals (intervals
with highest density).

\paragraph{$L_{F}(p)$ and $G_{F}$.}

Let us introduce the cumulated sum of the $n\le N$ \emph{smallest}
absolute errors
\begin{equation}
S_{n}=\sum_{i=1}^{n}\varepsilon_{[i]}
\end{equation}
where $\varepsilon_{[i]}$ is the $i^{th}$ order statistic (\emph{i.e.},
the value with rank $i$ ) of the sample of absolute errors. For consistency,
one sets $S_{0}=0$. 

The Lorenz curve is estimated as
\begin{equation}
L_{F}(p)=\frac{S_{p\times N}}{S_{N}}\label{eq:lorenz}
\end{equation}
where $0\le p\le1$. Note that, due to the use of finite samples,
$p$ takes its values in $\left\{ i/N\right\} _{i=0}^{N}$.

Using a fast sorting of the sample of absolute errors, an efficient
estimation of $G_{F}$ uses the formula \cite{Damgaard2000,Glasser1962,Dixon1987,ineq}
\begin{equation}
G_{F}=\frac{\sum_{i=1}^{N}(2i-N-1)\varepsilon_{[i]}}{N\sum_{i=1}^{N}\varepsilon_{[i]}}
\end{equation}
A slower, but equivalent expression in terms of mean values is \cite{Eliazar2010}
\begin{equation}
G_{F}=\frac{1}{MUE}<\max\{0,\varepsilon_{1}-\varepsilon_{2}\}>
\end{equation}
where $\varepsilon_{1}$ and $\varepsilon_{2}$ are two elements of
$\varXi$ and the mean is taken on all pairs. 

\paragraph{$\beta_{GM}$ and $\kappa_{CS}$.}

For skewness and kurtosis, Eq.\,\ref{eq:skewgm} and Eq.\,\ref{eq:kurtcs}
are applied directly, with the robust method to estimate quantiles
due to Harrell and Davis \cite{Harrell1982,Wilcox2012,Pernot2020}. 

\paragraph{Uncertainty.}

Uncertainty on any statistic $X$, noted $u(X)$, is estimated by
bootstrapping \cite{Efron1979} with 1000 samples. Note that there
is a known risk of underestimation of $G_{F}$ for small datasets
($N<100$) \cite{Dixon1987}.

\subsection{Implementation}

All calculations have been made in the\texttt{ R} language \cite{RTeam2019},
using several packages, notably for the Gini coefficient (package
\texttt{ineq} \cite{ineq}), the HRM estimator (package \texttt{genefilter}
\cite{genefilter}) and bootstrapping \texttt{(}package\texttt{ boot}
\cite{boot}). 

The Gini coefficient, Lorenz curves, $G_{MCF}$, $\beta_{GM}$, $\kappa_{CS}$
and mode estimator have been implemented in the freely available R
\cite{RTeam2019} package \texttt{ErrViewLib (v1.3,} \url{https://doi.org/10.5281/zenodo.3628475}).
The datasets can be analyzed with the \texttt{ErrView} graphical interface
(source: \url{https://doi.org/10.5281/zenodo.3628489}; web interface:
\url{http://upsa.shinyapps.io/ErrView}).

\section{Datasets}

\subsection{Model datasets}

Before applying the Gini coefficient to literature datasets, one explores
its properties on error sets generated from distributions with controlled
properties: uniform, normal, Student's-$t$, lognormal \cite{Evans2000}
and $g$-and-$h$ \cite{Hoaglin1985,Wilcox2012,Pernot2020}. 

It is important to note that we explore only distributions with a
single dominant, more or less structured peak, such as the ones encountered
in most computational chemistry error datasets. In the list of analytical
distributions above, the uniform is an exception because of its undefined
mode. We use it as an extreme case of single peaked, continuous distribution,
with negative excess kurtosis. 

Besides, it is easy to design multi-peaked distributions for which
our conclusions on the Gini coefficient would not be valid. In fact,
none of the usual summary statistics (MUE, MSE, skewness, kurtosis...)
would describe properly such distributions. 

\subsection{Literature datasets}

The statistical tools described above are applied to datasets gathered
in the computational chemistry literature. These are summarized in
Table\,\ref{tab:Case-studies}, and strongly overlap with those studied
in more details in a previous article \cite{Pernot2020a}, from which
we removed small datasets ($N<100$). The statistics, empirical cumulative
density functions and Lorenz curves corresponding to these datasets
are provided as Supplementary Information. 
\begin{table}[!t]
\noindent \begin{centering}
\begin{tabular}{llccl}
\hline 
Case & Property & $N$ & $K$ & Source\tabularnewline
\hline 
BOR2019 & Band gaps (eV) & 471 & 15 & \cite{Borlido2019}\tabularnewline
NAR2019 & Enthalpies of formation (kcal/mol) & 469 & 4 & \cite{Narayanan2019}\tabularnewline
PER2018 & Intensive atomization energies (kcal/mol) & 222 & 9 & \cite{Pernot2018}\tabularnewline
SCH2018 & Adsorption energies (eV) & 195 & 7 & \cite{Schmidt2018}\tabularnewline
THA2015 & Polarizability (relative errors, in \%) & 135 & 7 & \cite{Thakkar2015}\tabularnewline
WU2015 & Polarizability (relative errors, in \%) & 145 & 36 & \cite{Wu2015b}\tabularnewline
ZAS2019 & Effective atomization energies (kcal/mol) & 6211 & 3 & \cite{Zaspel2019}\tabularnewline
ZHA2018 & Solid formation enthalpies (kcal/mol) & 196 & 2 & \cite{Zhang2018}\tabularnewline
\hline 
\end{tabular}
\par\end{centering}
\caption{\label{tab:Case-studies}Literature datasets: $N$ is the number of
systems in the dataset and $K$ is the number of methods.}
\end{table}

\section{Applications\label{sec:Applications}}

In its usual application fields, the Gini coefficient is applied to
distributions with positive support. Our application to computational
chemistry error sets involves the intermediate folding operation,
which is not reversible. In a first part, we show how this limits
the information on the errors distribution that can be inferred from
the Gini coefficient. To relieve this difficulty, we propose a mode-centering
operation before folding, which better preserves some of the tail
properties of the original distributions. The Gini coefficient is
then compared to other tails statistics, notably at the level of statistical
uncertainty. 

\subsection{Gini coefficient \emph{vs.} bias\label{subsec:Gini-coefficient-vs.}}

\begin{figure}[!t]
\noindent \begin{centering}
\includegraphics[viewport=0bp 0bp 2400bp 1200bp,width=0.99\textwidth]{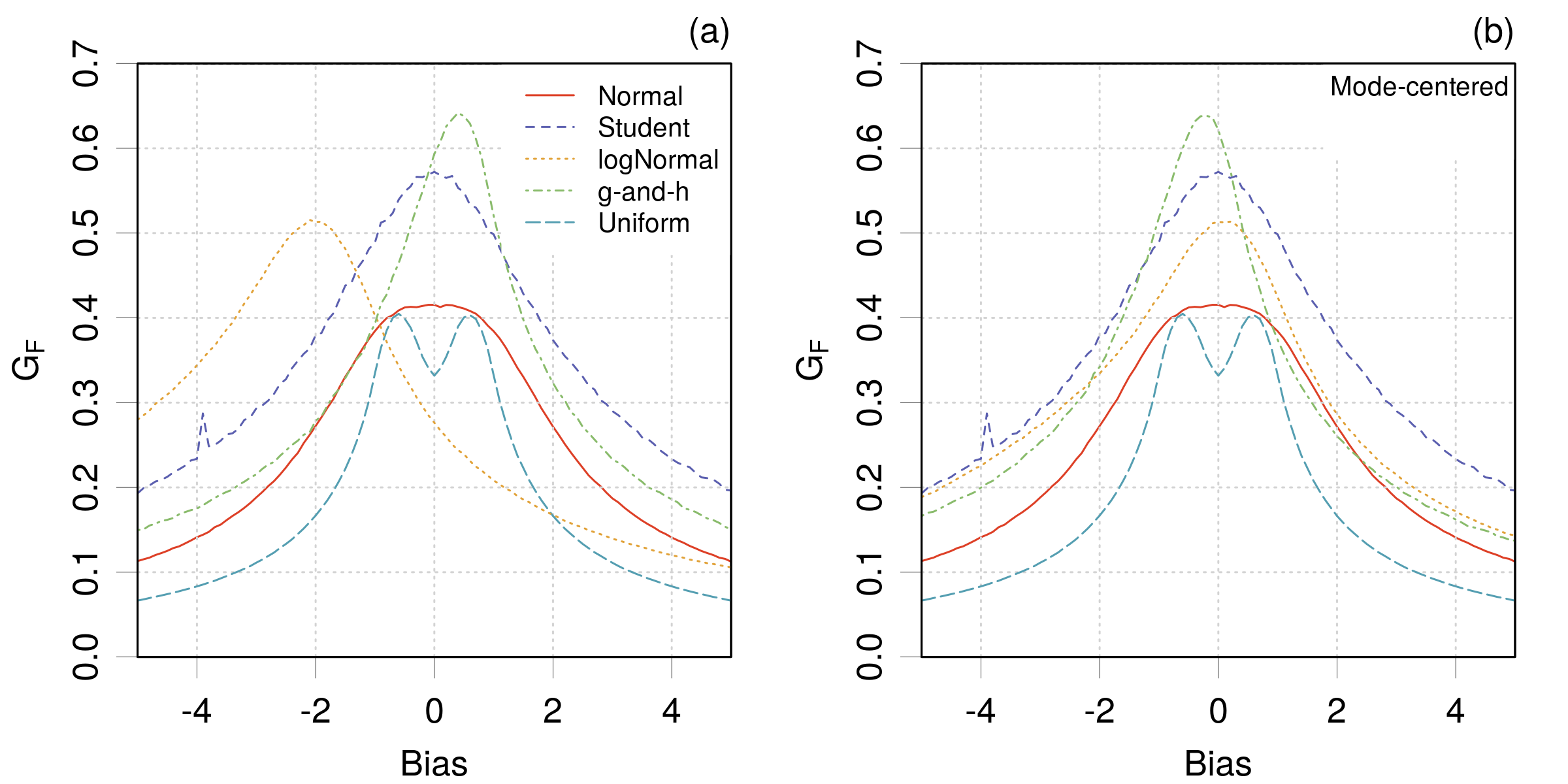}
\par\end{centering}
\caption{\label{fig:Gini-vs-bias}Variation of $G_{F}$ with a bias value added
to several model distributions: (a) no centering applied; (b) mode-centering
applied before adding bias (Uniform, Student and Normal are unchanged).}
\end{figure}

The link between the Gini coefficient and the coefficient of variation
(Eq.\,\ref{eq:GcsCv}) tells us that, for a normal distribution of
given standard deviation, an decreasing bias should result in increasing
values of $G_{F}$. At some point, this relation is broken by the
folding operation: as noted earlier, for a centered normal distribution,
one has $G_{F}=0.41$, for which Eq.\,\ref{eq:GcsCv} does not hold.
The dependence between a distribution shift and $G_{F}$ has been
plotted in Fig.\,\ref{fig:Gini-vs-bias}(a) for uniform $U(-1,1),$
normal $N(0,1)$, Student's-$t(\nu=2)$, lognormal $LN(1,0.5)$ and
$g$-and-$h$ $GH(g=1,h=0)$ distributions. 

The zero-centered unimodal symmetric distributions (normal and Student's-$t$)
have their maximal $G_{F}$ value when the bias is null. The $G_{F}$
curve for the uniform distribution reaches $G_{FU}=1/3$ when the
distribution is centered or shifted by $\pm1$. For intermediate absolute
values of the bias, the folded distribution is not uniform and presents
higher values of $G_{F}$. The value decreases when the bias is large
enough to exclude zero from the range of non-null densities. Having
heavier tails, the Student's-$t$ distribution has a larger Gini coefficient
than the normal. Any added bias leads to a decrease of $G_{F}$. 

The decay curves are symmetric with respect to the sign of the bias.
This is no longer the case for asymmetric distributions (lognormal
and $g$-and-$h$), for which the peak is reached for non-null values
of the bias and the decay curves are non-symmetric. 

In Fig.\,\ref{fig:Gini-vs-bias}(b) we plotted similar curves for
distributions centered on their mode before adding a bias (the symmetric
distributions have been left unchanged). This shows that the maximal
value of $G_{F}$ is reached when the mode of the distribution is
at, or near, the origin. This assertion is validated in the next section
(Sect.\,\ref{subsec:Does-mode-centering-maximize}).

Another important point illustrated by these curves is that the level
of information that can be recovered from $G_{F}$ is not uniform
over the range of $G_{F}$ values. For instance, $G_{F}=0.41$ might
as well occur for a centered normal distribution as for positively
or negatively biased Student's-$t$, lognormal of $g$-and-$h$ distributions,
whereas values above 0.41 exclude normal and uniform distributions.
Within the restrictions on the distributions shapes we considered
above, one might infer that a value smaller than $G_{FU}=1/3$ is
likely to reveal a biased error distribution, while a value above
$G_{FN}=0.41$ is likely to signal distributions with one or two extended
tail(s) (compared to normal tails), and with a possible bias. Between
these bounds lies a blind zone where compensation between bias and
shape factors prevent any inference on either of them.

\subsection{Does mode-centering maximize $G_{F}$ ?\label{subsec:Does-mode-centering-maximize}}

In order to avoid the blind zone effect observed above and be able
to characterize the shape of a distribution from its $G_{F}$ value,
mode-centering seems to offer an interesting way to relieve the bias/shape
compensation. Mode-centering a distribution results in a folded distribution
where both tails overlap and mix, but it ensures that the contribution
of the most-extended tail will prevail.\textcolor{orange}{{} }In absence
of mode-centering, when a biased distribution has a large tail encompassing
zero, the folding around zero might considerably reduce this tail. 

The assertion that mode-centering maximizes $G_{F}$ is tested here
by comparing the results for mode-centering with those obtained by
explicitly maximizing the Gini coefficient with respect to a bias
value. We define $b_{max}$ as the value of the bias which maximizes
$G_{F}$ 
\begin{equation}
b_{max}=\max_{b}G_{F}(|E-b|)
\end{equation}
 and note $G_{Fmax}=G_{F}(|E-b_{max}|)$. This equation is solved
numerically by the Nelder and Mead optimizer \cite{Nelder1965}. 

The values of $b_{max}$ and $G_{Fmax}$ were computed for the literature
datasets and compared to the mode $m(E)$ and $G_{MCF}$ respectively,
through $z$-scores 
\begin{equation}
z_{b}=\frac{m(E)-b_{max}}{\sqrt{u(m(E))^{2}+u(b_{max})^{2}}}
\end{equation}
 and 
\begin{equation}
z_{G}=\frac{G_{MCF}-G_{Fmax}}{\sqrt{u(G_{MCF})^{2}+u(G_{Fmax})^{2}}}
\end{equation}
where the uncertainties are estimated by bootstrap. 

In the hypothesis of a normal distribution of $z$-scores, a test
threshold of 2 is generally chosen for the absolute value of the $z$-score
\cite{Kacker2010}. For absolute values above 2, there is less than
5 percent of probability that the difference is due to sampling effects.
For values below, one does not reject the hypothesis that the compared
values are equal \cite{Pernot2020}. 

\begin{figure}[!t]
\noindent \begin{centering}
\includegraphics[viewport=0bp 0bp 2400bp 1200bp,width=0.95\textwidth]{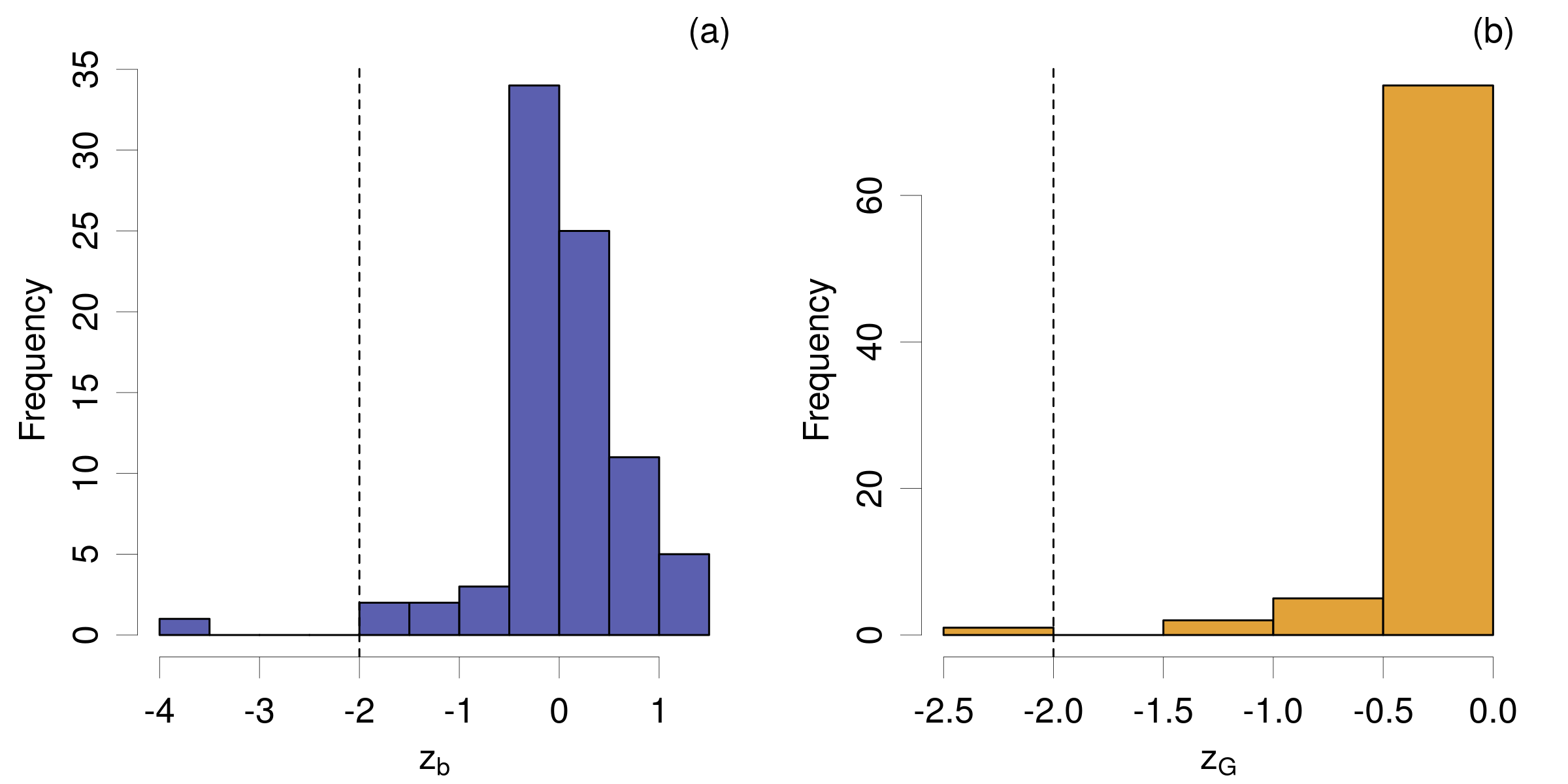}
\par\end{centering}
\caption{\label{fig:mode-vs-bmax}Histograms of $z$-scores for (a) the position
of the mode versus the $G_{F}$ maximizer; and (b) the values of the
corresponding Gini coefficients.}
\end{figure}

Histograms for the $z$-scores are shown in Fig.\,\ref{fig:mode-vs-bmax}.
At the exception of one point, the absolute value of all $z$-score
values are smaller than 2, and we have therefore no reason to reject
the hypothesis that these values are equal considering their uncertainty.
The outlying point, with $z_{b}=-3.8$ and $z_{G}=-2.5$ corresponds
to the MP2 method in dataset ZAS2019, which has a practically normal
distribution \cite{Pernot2020b}. One has $G_{MCF}=0.418(6)$ and
$G_{Fmax}=0.436(3)$ for a distance of 1.07 between the mode and $b_{max}$,
to be compared to the standard deviation of the distribution, 1.7.
As seen in Fig.\,\ref{fig:Gini-vs-bias}(a), there is a flat area
near the top of the $G_{F}$ curve as a function of bias for a normal
distribution: very small deviations from a perfect normal distribution
(as hinted to by the value of $G_{Fmax}$ being larger than 0.41)
can deviate the optimal point over a wide range. 

For all practical purposes in the present study, one can therefore
estimate that mode-centering maximizes the value of the Gini coefficient,
at a fraction of the computing cost for the search for $b_{max}$.
We further note that for distributions having distinct mean and mode,
centering on the mean would not maximize $G_{F}$ and therefore preserve
some of the ambiguity due to bias/shape compensation. 

\subsection{$G_{MCF}$ versus $G_{F}$\label{subsec:Mode-centering}}

\begin{figure}[!t]
\noindent \begin{centering}
\includegraphics[viewport=0bp 0bp 2400bp 2400bp,width=0.95\textwidth]{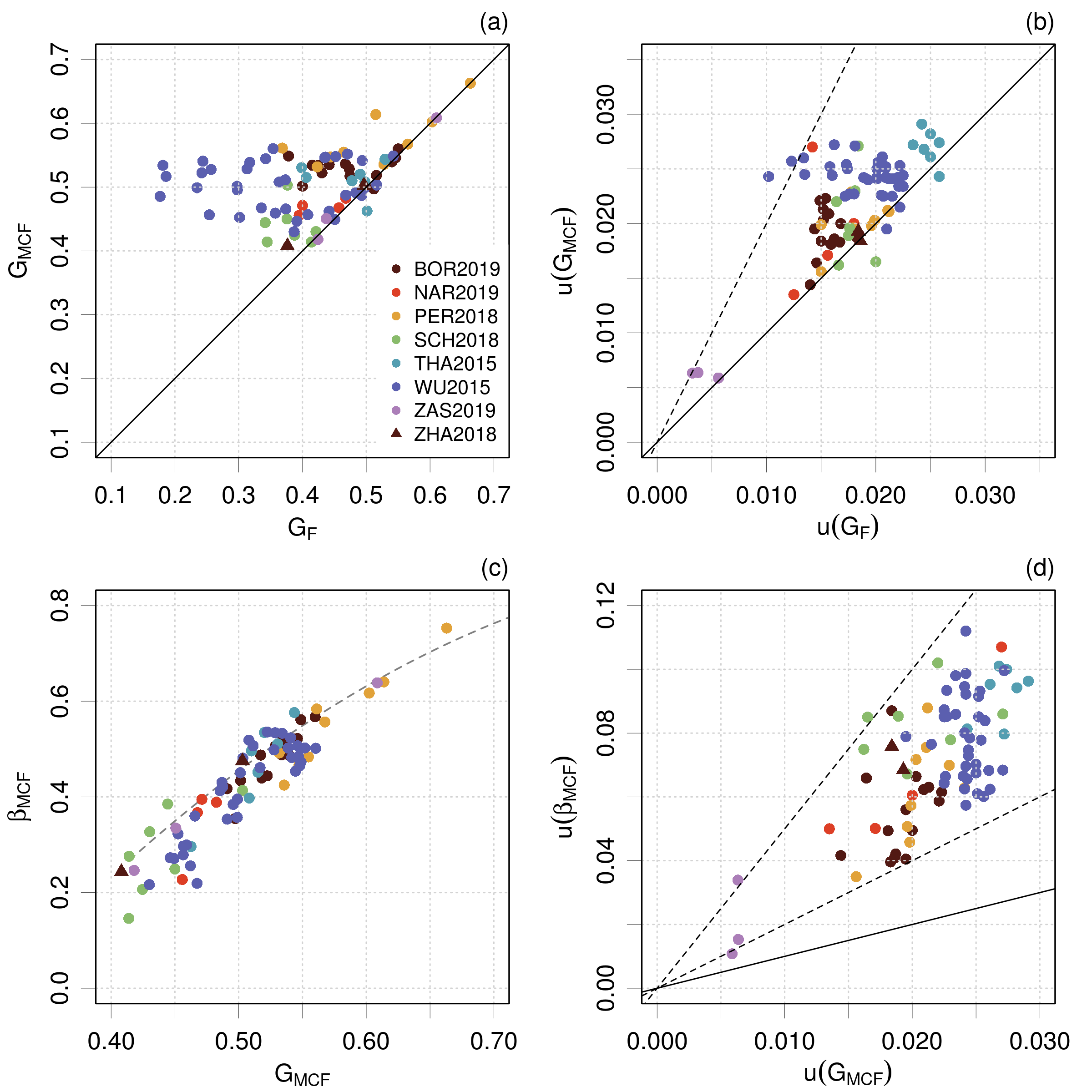}
\par\end{centering}
\caption{\label{fig:GF-vs-GMCF}Comparison of $G_{MCF}$ with other statistics:
(a) correlation of $G_{F}$ and $G_{MCF}$ for the literature datasets,
and (b) comparison of their uncertainties (the dashed line has a slope
of 2); (c) correlation of $G_{MCF}$ and $\beta_{MCF}$ for the literature
datasets (points) and a series of large samples ($N=10^{6}$) of Student's-$t$
and $g$-and-$h$ distributions with a range of shape parameters (dashed
line), and (d) comparison of their uncertainties (the points are bracketed
by lines of slope 2 and 5).}
\end{figure}

We note the Gini coefficient of mode-centered folded distributions
$G_{MCF}$. Fig.\,\ref{fig:GF-vs-GMCF}(a) displays $G_{MCF}$ versus
$G_{F}$ for the literature datasets. It is clear that mode-centering
increases all $G$ values, \emph{i.e.} $G_{MCF}\geq G_{F}$ for all
datasets, within the estimation uncertainties. The $G$-scale is now
reduced to values above 0.4, in conformity with our interpretation
that all values below $G_{FU}=1/3$ were due to bias. 

The uncertainties are reported in Fig.\,\ref{fig:GF-vs-GMCF}(b),
showing that for some datasets the uncertainty on $G_{MCF}$ is larger
than the uncertainty on $G_{F}$, up to a factor two. This extra uncertainty
is due to the uncertainty on the mode value. We note also a size effect,
the smallest datasets (THA2015, $N=145$) having the largest uncertainty,
and the largest dataset (ZAS2019, $N=6211)$, the smallest one. 

\subsection{$G_{MCF}$ as a shape statistic \label{subsec:GMCF-as-a-shape-stat}}

In parallel with the Gini coefficient, the skewness of the distribution
has also been considered as an estimator of inequality \cite{Bendel1989}.
One is interested here in comparing $G_{MCF}$ with $\beta_{MCF}$,
which is the skewness $\beta_{GM}$ (Eq.\,\ref{eq:skewgm}) of the
mode-centered folded distribution. 

The values for our selection of literature datasets is shown in Fig.\,\ref{fig:GF-vs-GMCF}(c).
There is an excellent correlation between those statistics, considering
the uncertainties reported in Fig.\,\ref{fig:GF-vs-GMCF}(d). Using
model datasets of large size ($N=10^{6}$) for Student's-$t$ and
$g$-and-$h$ distributions with a range of shape parameters, one
observes a nearly perfect non-linear correlation (dashed line, resulting
from a quadratic fit of the sampled values). The dispersion of the
points for the literature datasets about this curve is mostly due
to statistical uncertainty (the points for the largest dataset (ZAS2019,
$N=6211$) are very close to the curve). It is important to note that
the uncertainty on $G_{MCF}$ is a factor two to five smaller than
the uncertainty on $\beta_{MCF}$, and therefore performs better for
smaller datasets.

\begin{figure}[!t]
\noindent \begin{centering}
\includegraphics[viewport=0bp 0bp 2400bp 1200bp,width=0.95\textwidth]{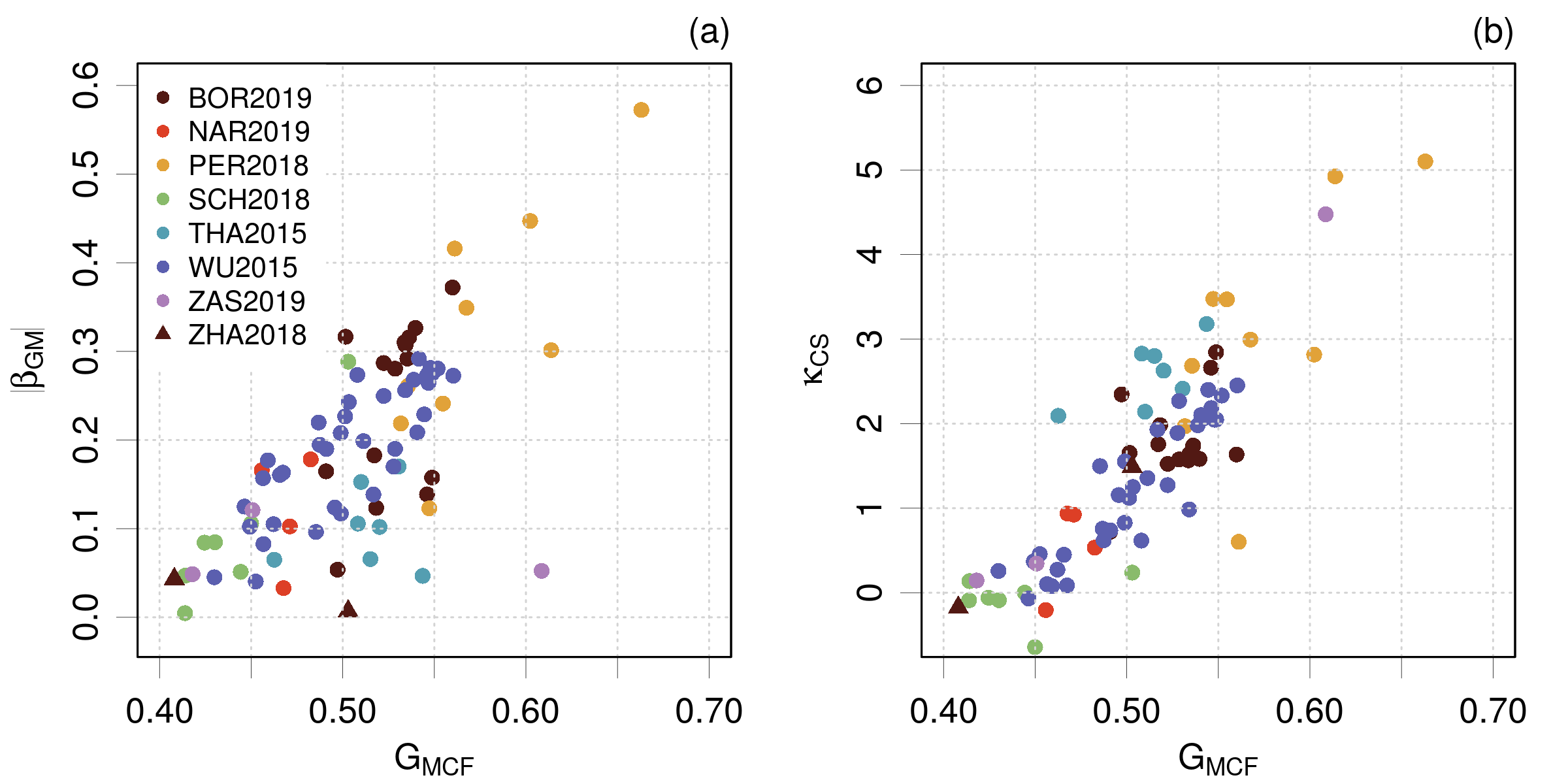}
\par\end{centering}
\caption{\label{fig:GMCF-vs-skew}Comparison of $G_{MCF}$ with shape statistics
of error distributions: (a) absolute value of skewness $|\beta_{GM}|$;
(b) excess kurtosis $\kappa_{CS}$.}
\end{figure}

We can therefore conclude that $G_{MCF}$ is apt at estimating the
heaviness of the errors distribution tail after mode-centering and
folding. In order to appreciate the information about the \emph{signed}
error distribution that can be extracted from $G_{MCF}$, we plotted
it against the skewness $\beta_{GM}$ and excess kurtosis $\kappa_{CS}$
(Fig.\,\ref{fig:GMCF-vs-skew}). 

Considering skewness (Fig.\,\ref{fig:GMCF-vs-skew}(a)), all the
points seem to lie within an angular sector, indicating that distributions
with large skewness have necessarily large $G_{MCF}$ values. For
instance, if the absolute value of the skewness is above 0.3, $G_{MCF}$
is larger than 0.5. Reciprocally, $G_{MCF}$ provides only an upper
limit to $|\beta_{GM}|$ (for $G_{MCF}=0.55$, the absolute value
of the skewness cannot be above 0.4). For kurtosis (Fig.\,\ref{fig:GMCF-vs-skew}(b)),
there is a lax positive trend between both statistics, and, globally,
large values of $G_{MCF}$ are associated with large excess kurtosis,
which might be due to heavy tails or outliers. In both graphs, values
of $G_{MCF}$ below 0.45 are associated with low skewness \emph{and}
excess kurtosis. Although information is lost because of folding,
$G_{MCF}$ can still provide some information about the shape of the
distribution of signed errors, and notably about the kurtosis.

Let us consider a few examples to illustrate this point. We see in
Fig.\,\ref{fig:GMCF-vs-skew} that most points fall between 0.4 and
0.55, but a few methods reach higher values. The largest $G_{MCF}$
value in this study is 0.66 (orange dot) for method CAM-B3LYP in the
PER2018 set \cite{Pernot2018} (\emph{cf.} Table\,\ref{tab:statsLit}).
This corresponds to large values of both $\beta_{GM}$ and $\kappa_{CS}$.
The authors discussed how this DFA is in the head group of two methods
with similar MUE values, but does not minimize the risk of large errors
(Sect.\,II.A \cite{Pernot2018}). From the same set, B3LYP has the
second largest $G_{MCF}$ value (0.61) and presents the same tail
features than CAM-B3LYP. The third largest $G_{MCF}$ value (0.61,
violet dot) belongs to the ZAS2019 dataset, and it presents a null
skewness and a large kurtosis. An in-depth study has been published
for this case \cite{Pernot2020b}, where the errors distribution for
the SLATM-L2 method was shown to have large tails, despite having
the smallest MUE among the compared methods. In the same set, the
MP2 method has a practically normal errors distribution and can be
found in the lower part of the Gini scale (0.42). This analysis can
be repeated for the ten methods with largest $G_{MCF}$ values (Table\,\ref{tab:statsLit}),
showing that large $G_{MCF}$ values point indeed towards error sets
with high kurtosis and/or skewness. 
\begin{table}[ht]
\noindent \centering{}%
\begin{tabular}{llr@{\extracolsep{0pt}.}lr@{\extracolsep{0pt}.}lr@{\extracolsep{0pt}.}l}
\hline 
Dataset  & Methods  & \multicolumn{2}{c}{$G_{MCF}$} & \multicolumn{2}{c}{$\beta_{GM}$} & \multicolumn{2}{c}{$\kappa_{CS}$}\tabularnewline
\hline 
PER2018  & CAM-B3LYP  & 0&663(16)  & 0&572(57)  & 5&1(1.3) \tabularnewline
PER2018  & B3LYP  & 0&614(20)  & 0&301(74)  & 4&93(96) \tabularnewline
ZAS2019  & SLATM\_L2  & 0&6086(59)  & 0&052(22)  & 4&48(25) \tabularnewline
PER2018  & LC-$\omega$PBE  & 0&602(20)  & 0&447(64)  & 2&82(82) \tabularnewline
PER2018  & PBE0  & 0&568(24)  & 0&349(63)  & 2&99(86) \tabularnewline
PER2018  & BH\&HLYP  & 0&561(20)  & 0&416(52)  & 0&60(48) \tabularnewline
WU2015  & $\tau$HCTHhyb  & 0&560(24)  & -0&273(84)  & 2&45(83) \tabularnewline
BOR2019  & HLE16 + SOC  & 0&560(19)  & 0&372(43)  & 1&64(47) \tabularnewline
PER2018  & BLYP  & 0&555(20)  & -0&241(67)  & 3&47(78) \tabularnewline
WU2015  & B97-2  & 0&552(22)  & -0&281(80)  & 2&33(77) \tabularnewline
\hline 
\end{tabular}\caption{\label{tab:statsLit}The ten methods with the largest $G_{MCF}$ values,
and the corresponding skewness and kurtosis.}
\end{table}

\subsection{Application of $G_{MCF}$ to ranking}

To evaluate the interest of $G_{MCF}$ in a practical scenario, one
might consider it as an alert mechanism to complement a MUE ranking.
But a question remains: ``What is the threshold for $G_{MCF}$ one
should use to detect problematic error distributions ?'' We have
seen above that the ten largest $G_{MCF}$ values, above 0.55, point
to distributions with notable tails. We propose for the present study
to adopt an alert value based on the median of the $G_{MCF}$ values
for our full dataset (0.51) and round it to 0.5. This might be reevaluated
when more data are analyzed. Using this threshold, one might flag
distributions suspected of having tails unsuitable for reliable predictions.
\begin{figure}[!t]
\noindent \begin{centering}
\includegraphics[width=0.95\textwidth]{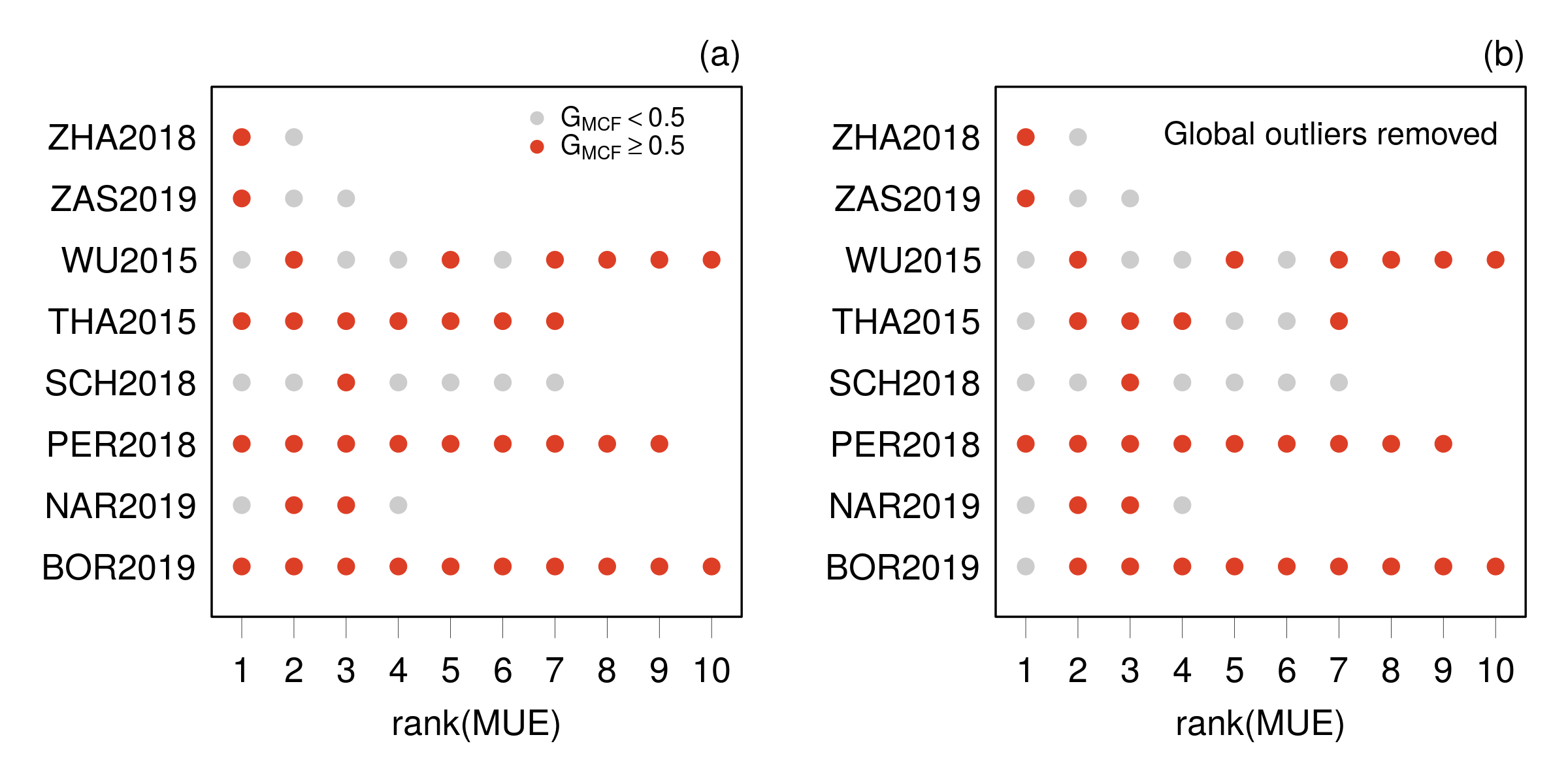}
\par\end{centering}
\caption{\label{fig:Gini-vs-CV-Rank}Flagging of large $G_{MCF}$ values after
mode-centering of at most 10 lowest MUE-ranked methods in each dataset. }
\end{figure}

Fig.\,\ref{fig:Gini-vs-CV-Rank}(a) shows, for each dataset, the
flagging of the methods with best ranking. If one considers the first
rank, five methods are flagged, but it is striking that three datasets
have all their 10 lowest MUE-ranked methods flagged (BOR2019, PER2018
and THA2015). For BOR2019, all methods present some excess kurtosis
and variable levels of skewness. We can relate this to an increasing
trend of the errors with the bandgap value \cite{Pernot2020a}. In
the case of PER2018, the error distributions present also large skewness
and kurtosis, that can be associated with the chemical heterogeneity
of the dataset \cite{Pernot2018}. For THA2015, it was noted previously
\cite{Thakkar2015,Pernot2020a} that some experimental reference data
with large measurement uncertainty could not be reproduced by any
method in the studied set. These outliers contribute to the tails
of all the error distributions (so-called \emph{global} outliers)
and affect $G_{MCF}$ values. Note that, more generally, reference
data are not necessarily the origin of global outliers, as a missing
physical contribution in the tested methods could produce similar
effects. 

To explore the role of global outliers, Fig.\,\ref{fig:Gini-vs-CV-Rank}(b)
reports the same analysis after search and removal of global outliers,
defined as systems lying out of the $[q(0.025),q(0.975)]$ interval
for \emph{all methods} of a dataset. The removal of 6 systems affects
strongly the case THA2015, confirming the previous analysis. The results
for BOR2019 are mostly unchanged, except for the best MUE-ranked method
(mBJ) which benefits from the removal of a single global outlier.
No effect is observed for the PER2018 dataset, confirming the intrinsic
heavy-tailed shape of these heterogeneous atomization energy error
sets \cite{Pernot2018}.

The other datasets with leading $G_{MCF}$-flagged methods are ZAS2019
and ZHA2018. The former has already been discussed (Sect.\,\ref{subsec:GMCF-as-a-shape-stat}),
and the removal of several global outliers has no impact. In the case
ZHA2018, the first MUE-ranked method is SCAN, which has a $G_{MCF}$
value just above the threshold (0.503), presents no skewness and a
slight level of kurtosis. Removal of 4 global outliers does not improve
the shape of the errors distribution. However, the errors on the formation
enthalpies present a linear trend as a function of the calculated
values. Correcting this trend \cite{Pernot2015} improves slightly
the performance and the shape of the SCAN distribution, but most notably
of the PBE distribution, which performances get indistinguishable
from those of SCAN (Table\,\ref{tab:statsZHA}). 
\begin{table}[t]
\noindent \centering{}%
\begin{tabular}{lr@{\extracolsep{0pt}.}lr@{\extracolsep{0pt}.}lr@{\extracolsep{0pt}.}lr@{\extracolsep{0pt}.}lr@{\extracolsep{0pt}.}lr@{\extracolsep{0pt}.}l}
\hline 
Methods  & \multicolumn{2}{c}{MUE} & \multicolumn{2}{c}{MSE } & \multicolumn{2}{c}{$Q_{95}$} & \multicolumn{2}{c}{$\beta_{GM}$} & \multicolumn{2}{c}{$\kappa_{CS}$} & \multicolumn{2}{c}{$G_{MCF}$}\tabularnewline
 & \multicolumn{2}{c}{(kcal/mol)} & \multicolumn{2}{c}{(kcal/mol)} & \multicolumn{2}{c}{(kcal/mol)} & \multicolumn{2}{c}{} & \multicolumn{2}{c}{} & \multicolumn{2}{c}{}\tabularnewline
\hline 
PBE  & 0&2106(98)  & -0&205(10)  & 0&467(34)  & -0&043(64)  & -0&17(28)  & 0&408(18) \tabularnewline
SCAN  & 0&1024(69)  & -0&0165(97)  & 0&291(21)  & -0&007(67)  & 1&49(53)  & 0&503(19) \tabularnewline
lc-PBE  & 0&0923(61)  & 0&0 & 0&287(38)  & -0&138(71)  & 1&02(43)  & 0&479(24) \tabularnewline
lc-SCAN  & 0&0917(63)  & 0&0 & 0&276(26)  & -0&082(67)  & 1&27(41)  & 0&475(20) \tabularnewline
\hline 
\end{tabular}\caption{\label{tab:statsZHA}Statistics for the methods of the ZHA2018 dataset,
before and after linear correction ('lc-' prefix). }
\end{table}

\subsection{Limits of the $G_{MCF}$ coefficient}

\begin{figure}[!t]
\noindent \begin{centering}
\includegraphics[viewport=0bp 0bp 1200bp 1200bp,height=8cm]{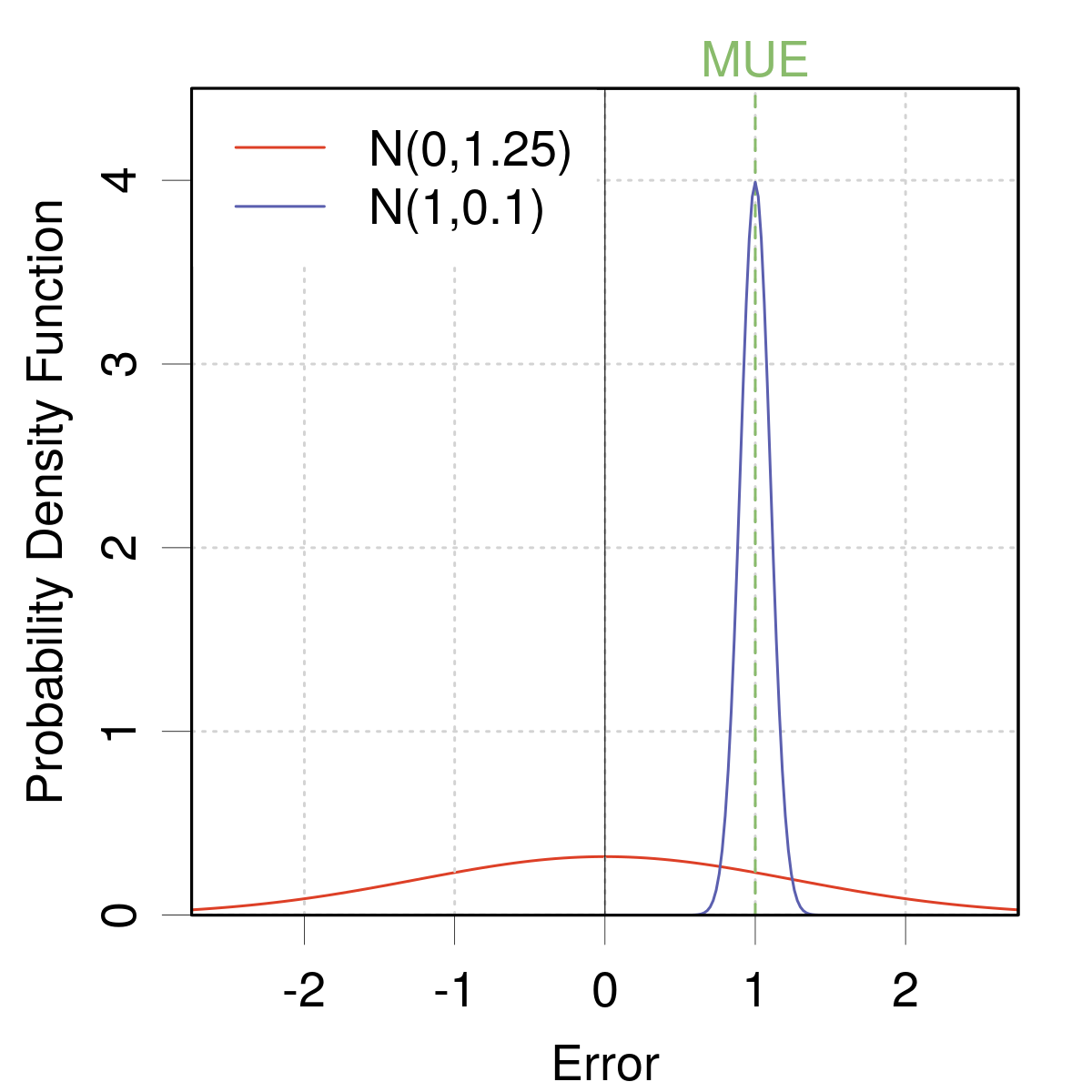}
\par\end{centering}
\caption{\label{fig:ambiguity}Example of different error distributions having
the same MUE (1.0) and offering contradictory results for some tail
statistics. The probability to have \emph{absolute} errors larger
than 1.0 is $P_{1}=0.50$ for the blue curve and 0.42 for the red
curve, hiding the fact that the red distribution contains much worse
results that the blue one. In this case, the problem is solved by
the values of $Q_{95}$, giving 1.16 for the blue curve, and 2.46
for the red one. Shape statistics, such as the kurtosis, would not
enable to discriminate between both normal distributions. }
\end{figure}

We have shown above that $G_{MCF}$ might be a useful complement to
the usual ranking statistics, in order to detect error distributions
with shapes that might reveal problem in prediction reliability. However
there remains cases where the $G_{MCF}$ index is insufficient to
reveal underlying problems. Fig.\,\ref{fig:ambiguity} proposes a
scenario of two normal distributions ($G_{MCF}=0.41$) with the same
value of the MUE (1.0), and yet very different risks of large prediction
errors. This is clearly a case showing that a quantile-based statistic,
such as $Q_{95}$, is an essential complement to the MUE. 

\section{Conclusion}

The Gini coefficient presents an interesting addition to the computational
chemistry benchmarking statistical toolbox. We focused here on its
properties in relation with features of the error distributions, such
as bias and shape (skewness and kurtosis). The interest of the Gini
coefficient is that it correlates with these features, and offers
a one-number summary. This is also one of its weaknesses, as there
is no unique mapping from the Gini coefficient to these features. 

To unscramble this situation, we propose to use the Gini coefficient
of the mode-centered distributions, $G_{MCF}$, which offers a simpler
to interpret, shape-based, measure of tailedness. Large $G_{MCF}$
values, \emph{e.g.} above 0.5, alert us about large tails that might
be due to large skewness, kurtosis and/or the presence of outliers.
For high ranking methods, this is an incentive to inspect closely
the error distributions and check if the selected methods might have
problems of reliability in their predictions. It might then be worth
to investigate if the distorted shape of the distribution is due to
systematic trends in the errors, as they can often be corrected by
simple linear transformations \cite{Lejaeghere2014,Pernot2015,Lejaeghere2016,Proppe2016,Proppe2017}.
The impact of such corrections on the shape of error distributions
is the prospect of further studies.

\section*{Supplementary Information}

Statistics, ECDFs and Lorenz curves for the literature datasets are
openly available at the following URL: \url{https://doi.org/10.5281/zenodo.4333217}

\section*{Data availability statement }

The data and codes that support the findings of this study are openly
available at the following URL: \url{https://doi.org/10.5281/zenodo.4333217} 

\section*{Dedication}

This work is dedicated to Ramon Carbó-Dorca for his 80th birthday.
It reflects his interest for cross-disciplinary aspects of science,
especially the role of mathematics in chemistry. 

\bibliographystyle{unsrturlPP}
\bibliography{NN}

\begin{thebibliography}{10}

\bibitem{Pernot2015}
P.~Pernot, B.~Civalleri, D.~Presti, and A.~Savin.
\newblock \href{http://dx.doi.org/10.1021/jp509980w}{Prediction uncertainty of
  density functional approximations for properties of crystals with cubic
  symmetry}.
\newblock {\em J. Phys. Chem. A}, 119:5288--5304, 2015.
\newblock \href {https://doi.org/10.1021/jp509980w}
  {\path{doi:10.1021/jp509980w}}.

\bibitem{Pernot2018}
P.~Pernot and A.~Savin.
\newblock \href{http://dx.doi.org/10.1063/1.5016248}{Probabilistic performance
  estimators for computational chemistry methods: the empirical cumulative
  distribution function of absolute errors}.
\newblock {\em J. Chem. Phys.}, 148:241707, 2018.
\newblock \href {https://doi.org/10.1063/1.5016248}
  {\path{doi:10.1063/1.5016248}}.

\bibitem{Pernot2020}
P.~Pernot and A.~Savin.
\newblock \href{http://dx.doi.org/10.1063/5.0006202}{Probabilistic performance
  estimators for computational chemistry methods: Systematic improvement
  probability and ranking probability matrix. {I. Theory}}.
\newblock {\em J. Chem. Phys.}, 152:164108, 2020.
\newblock \href {https://doi.org/10.1063/5.0006202}
  {\path{doi:10.1063/5.0006202}}.

\bibitem{Pernot2020a}
P.~Pernot and A.~Savin.
\newblock \href{http://dx.doi.org/10.1063/5.0006204}{Probabilistic performance
  estimators for computational chemistry methods: Systematic improvement
  probability and ranking probability matrix. {II. Applications}}.
\newblock {\em J. Chem. Phys.}, 152:164109, 2020.
\newblock \href {https://doi.org/10.1063/5.0006204}
  {\path{doi:10.1063/5.0006204}}.

\bibitem{Pernot2020b}
P.~Pernot, B.~Huang, and A.~Savin.
\newblock \href{http://dx.doi.org/10.1088/2632-2153/aba184}{Impact of
  non-normal error distributions on the benchmarking and ranking of {Quantum}
  {Machine} {Learning} models}.
\newblock {\em Mach. Learn.: Sci. Technol.}, 1:035011, 2020.
\newblock \href {https://doi.org/10.1088/2632-2153/aba184}
  {\path{doi:10.1088/2632-2153/aba184}}.

\bibitem{Bonato2011}
M.~Bonato.
\newblock \href{http://dx.doi.org/10.1016/j.frl.2010.12.001}{Robust estimation
  of skewness and kurtosis in distributions with infinite higher moments}.
\newblock {\em Finance Research Letters}, 8:77--87, 2011.
\newblock \href {https://doi.org/10.1016/j.frl.2010.12.001}
  {\path{doi:10.1016/j.frl.2010.12.001}}.

\bibitem{Lorenz1905}
M.~O. Lorenz.
\newblock \href{http://dx.doi.org/10.1080/15225437.1905.10503443}{Methods of
  measuring the concentration of wealth}.
\newblock {\em Publications of the American Statistical Association},
  9:209--219, 1905.
\newblock \href {https://doi.org/10.1080/15225437.1905.10503443}
  {\path{doi:10.1080/15225437.1905.10503443}}.

\bibitem{Gini1912}
C.~{Gini}.
\newblock {\em {Variabilit{\`a} e mutabilit{\`a}}}.
\newblock 1912.

\bibitem{Damgaard2000}
C.~Damgaard and J.~Weiner.
\newblock \href{http://dx.doi.org/10.2307/177185}{Describing inequality in
  plant size or fecundity}.
\newblock {\em Ecology}, 81:1139--1142, 2000.
\newblock \href {https://doi.org/10.2307/177185} {\path{doi:10.2307/177185}}.

\bibitem{Eliazar2010}
I.~I. Eliazar and I.~M. Sokolov.
\newblock \href{http://dx.doi.org/10.1016/j.physa.2009.08.006}{Measuring
  statistical heterogeneity: The {P}ietra index}.
\newblock {\em Physica A}, 389:117--125, 2010.
\newblock \href {https://doi.org/10.1016/j.physa.2009.08.006}
  {\path{doi:10.1016/j.physa.2009.08.006}}.

\bibitem{Bendel1989}
R.~B. Bendel, S.~S. Higgins, J.~E. Teberg, and D.~A. Pyke.
\newblock \href{http://dx.doi.org/10.1007/BF00379115}{Comparison of skewness
  coefficient, coefficient of variation, and {G}ini coefficient as inequality
  measures within populations}.
\newblock {\em Oecologia}, 78:394--400, 1989.
\newblock \href {https://doi.org/10.1007/BF00379115}
  {\path{doi:10.1007/BF00379115}}.

\bibitem{Florian2016}
M.~K. Florian, N.~Li, and M.~D. Gladders.
\newblock \href{http://dx.doi.org/10.3847/0004-637X/832/2/168}{The {G}ini
  coefficient as a morphological measurement of strongly lensed galaxies in the
  image plane}.
\newblock {\em Astrophys. J.}, 832:168, 2016.
\newblock \href {https://doi.org/10.3847/0004-637X/832/2/168}
  {\path{doi:10.3847/0004-637X/832/2/168}}.

\bibitem{Hurley2009}
N.~Hurley and S.~Rickard.
\newblock \href{http://dx.doi.org/10.1109/TIT.2009.2027527}{Comparing measures
  of sparsity}.
\newblock {\em IEEE Transactions on Information Theory}, 55:4723--4741, 2009.
\newblock \href {https://doi.org/10.1109/TIT.2009.2027527}
  {\path{doi:10.1109/TIT.2009.2027527}}.

\bibitem{Kleiber2005}
C.~Kleiber.
\newblock \href{http://dx.doi.org/10.17877/DE290R-14481}{The {Lorenz} curve in
  economics and econometrics}.
\newblock techreport, TU Dortmund, March 2005.
\newblock \href {https://doi.org/10.17877/DE290R-14481}
  {\path{doi:10.17877/DE290R-14481}}.

\bibitem{Dixon1987}
P.~M. Dixon, J.~Weiner, T.~Mitchell-Olds, and R.~Woodley.
\newblock \href{http://dx.doi.org/10.2307/1939238}{Bootstrapping the {G}ini
  coefficient of inequality}.
\newblock {\em Ecology}, 68:1548--1551, 1987.
\newblock \href {https://doi.org/10.2307/1939238} {\path{doi:10.2307/1939238}}.

\bibitem{Ruppert1987}
D.~Ruppert.
\newblock \href{http://dx.doi.org/10.2307/2684309}{What is kurtosis?: An
  influence function approach}.
\newblock {\em The American Statistician}, 41:1, 1987.
\newblock \href {https://doi.org/10.2307/2684309} {\path{doi:10.2307/2684309}}.

\bibitem{Groeneveld1984}
R.~A. Groeneveld and G.~Meeden.
\newblock \href{http://dx.doi.org/10.2307/2987742}{Measuring skewness and
  kurtosis}.
\newblock {\em The Statistician}, 33:391--399, 1984.
\newblock URL: \url{http://www.jstor.org/stable/2987742}, \href
  {https://doi.org/10.2307/2987742} {\path{doi:10.2307/2987742}}.

\bibitem{Suaray2015}
K.~Suaray.
\newblock \href{http://dx.doi.org/10.14419/ijasp.v3i2.5007}{On the asymptotic
  distribution of an alternative measure of kurtosis}.
\newblock {\em Int. J. Adv. Stat. Proba.}, 3:161--168, 2015.
\newblock \href {https://doi.org/10.14419/ijasp.v3i2.5007}
  {\path{doi:10.14419/ijasp.v3i2.5007}}.

\bibitem{Crow1967}
E.~L. Crow and M.~M. Siddiqui.
\newblock \href{http://dx.doi.org/10.2307/2283968}{Robust estimation of
  location}.
\newblock {\em Journal of the American Statistical Association}, 62:353--389,
  1967.
\newblock \href {https://doi.org/10.2307/2283968} {\path{doi:10.2307/2283968}}.

\bibitem{Bickel2002}
D.~R. Bickel.
\newblock \href{http://dx.doi.org/10.1016/S0167-9473(01)00057-3}{Robust
  estimators of the mode and skewness of continuous data}.
\newblock {\em Comput. Stat. Data Anal.}, 39:153--163, 2002.
\newblock \href {https://doi.org/10.1016/S0167-9473(01)00057-3}
  {\path{doi:10.1016/S0167-9473(01)00057-3}}.

\bibitem{Hedges2003}
S.~B. Hedges and P.~Shah.
\newblock \href{http://dx.doi.org/10.1186/1471-2105-4-31}{Comparison of mode
  estimation methods and application in molecular clock analysis}.
\newblock {\em BMC Bioinformatics}, 4:31, 2003.
\newblock \href {https://doi.org/10.1186/1471-2105-4-31}
  {\path{doi:10.1186/1471-2105-4-31}}.

\bibitem{Glasser1962}
G.~J. Glasser.
\newblock \href{http://dx.doi.org/10.1080/01621459.1962.10500553}{Variance
  formulas for the mean difference and coefficient of concentration}.
\newblock {\em J. Am. Stat. Assoc.}, 57:648--654, 1962.
\newblock \href {https://doi.org/10.1080/01621459.1962.10500553}
  {\path{doi:10.1080/01621459.1962.10500553}}.

\bibitem{ineq}
A.~Zeileis.
\newblock \href{https://CRAN.R-project.org/package=ineq}{{\em ineq: Measuring
  Inequality, Concentration, and Poverty}}, 2014.
\newblock R package version 0.2-13.
\newblock URL: \url{https://CRAN.R-project.org/package=ineq}.

\bibitem{Harrell1982}
F.~E. Harrell and C.~Davis.
\newblock \href{http://dx.doi.org/10.2307/2335999}{A new distribution-free
  quantile estimator}.
\newblock {\em Biometrika}, 69:635--640, 1982.
\newblock \href {https://doi.org/10.2307/2335999} {\path{doi:10.2307/2335999}}.

\bibitem{Wilcox2012}
R.~R. Wilcox and D.~M. Erceg-Hurn.
\newblock \href{http://dx.doi.org/10.1080/02664763.2012.724665}{Comparing two
  dependent groups via quantiles}.
\newblock {\em J. App. Stat.}, 39:2655--2664, 2012.
\newblock \href {https://doi.org/10.1080/02664763.2012.724665}
  {\path{doi:10.1080/02664763.2012.724665}}.

\bibitem{Efron1979}
B.~Efron.
\newblock \href{http://dx.doi.org/10.1214/aos/1176344552}{Bootstrap {Methods}:
  {Another} {Look} at the {Jackknife}}.
\newblock {\em Ann. Stat.}, 7(1):1--26, January 1979.
\newblock \href {https://doi.org/10.1214/aos/1176344552}
  {\path{doi:10.1214/aos/1176344552}}.

\bibitem{RTeam2019}
{R Core Team}.
\newblock \href{http://www.R-project.org/}{{\em {R}: {A} {L}anguage and
  {E}nvironment for {S}tatistical {C}omputing}}.
\newblock R Foundation for Statistical Computing, Vienna, Austria, 2019.
\newblock URL: \url{http://www.R-project.org/}.

\bibitem{genefilter}
R.~Gentleman, V.~Carey, W.~Huber, and F.~Hahne.
\newblock \href{http://dx.doi.org/10.18129/B9.bioc.genefilter}{{\em genefilter:
  methods for filtering genes from high-throughput experiments}}, 2019.
\newblock R package version 1.68.0.
\newblock \href {https://doi.org/10.18129/B9.bioc.genefilter}
  {\path{doi:10.18129/B9.bioc.genefilter}}.

\bibitem{boot}
A.~Canty and B.~D. Ripley.
\newblock {\em boot: Bootstrap R (S-Plus) Functions}, 2019.
\newblock R package version 1.3-22.

\bibitem{Evans2000}
M.~Evans, N.~Hastings, and B.~Peacock.
\newblock {\em Statistical Distributions}.
\newblock Wiley-Interscience, 3rd edition, 2000.

\bibitem{Hoaglin1985}
D.~C. Hoaglin.
\newblock {\em Exploring data tables, trends, and shapes}, chapter Summarizing
  shape numerically: The g-and-h distributions, pages 461--513.
\newblock Wiley, New York, 1985.

\bibitem{Borlido2019}
P.~Borlido, T.~Aull, A.~W. Huran, F.~Tran, M.~A. Marques, and S.~Botti.
\newblock \href{http://dx.doi.org/10.1021/acs.jctc.9b00322}{Large-scale
  benchmark of exchange--correlation functionals for the determination of
  electronic band gaps of solids}.
\newblock {\em J. Chem. Theory Comput.}, 15:5069--5079, 2019.
\newblock \href {https://doi.org/10.1021/acs.jctc.9b00322}
  {\path{doi:10.1021/acs.jctc.9b00322}}.

\bibitem{Narayanan2019}
B.~Narayanan, P.~C. Redfern, R.~S. Assary, and L.~A. Curtiss.
\newblock \href{http://dx.doi.org/10.1039/c9sc02834j}{Accurate quantum chemical
  energies for 133000 organic molecules}.
\newblock {\em Chem. Sci.}, 10:7449--7455, 2019.
\newblock \href {https://doi.org/10.1039/c9sc02834j}
  {\path{doi:10.1039/c9sc02834j}}.

\bibitem{Schmidt2018}
P.~S. Schmidt and K.~S. Thygesen.
\newblock \href{http://dx.doi.org/10.1021/acs.jpcc.7b12258}{Benchmark database
  of transition metal surface and adsorption energies from many-body
  perturbation theory}.
\newblock {\em J. Phys. Chem. C}, 122:4381--4390, 2018.
\newblock \href {https://doi.org/10.1021/acs.jpcc.7b12258}
  {\path{doi:10.1021/acs.jpcc.7b12258}}.

\bibitem{Thakkar2015}
A.~J. Thakkar and T.~Wu.
\newblock \href{http://dx.doi.org/10.1063/1.4932594}{How well do static
  electronic dipole polarizabilities from gas-phase experiments compare with
  density functional and {MP2} computations?}
\newblock {\em J. Chem. Phys.}, 143:144302, 2015.
\newblock \href {https://doi.org/10.1063/1.4932594}
  {\path{doi:10.1063/1.4932594}}.

\bibitem{Wu2015b}
T.~Wu, Y.~N. Kalugina, and A.~J. Thakkar.
\newblock \href{http://dx.doi.org/10.1016/j.cplett.2015.07.003}{Choosing a
  density functional for static molecular polarizabilities}.
\newblock {\em Chem. Phys. Lett.}, 635:257--261, 2015.
\newblock \href {https://doi.org/10.1016/j.cplett.2015.07.003}
  {\path{doi:10.1016/j.cplett.2015.07.003}}.

\bibitem{Zaspel2019}
P.~Zaspel, B.~Huang, H.~Harbrecht, and O.~A. von Lilienfeld.
\newblock \href{http://dx.doi.org/10.1021/acs.jctc.8b00832}{Boosting quantum
  machine learning models with a multilevel combination technique: Pople
  diagrams revisited}.
\newblock {\em J. Chem. Theory Comput.}, 15(3):1546--1559, 2019.
\newblock \href {https://doi.org/10.1021/acs.jctc.8b00832}
  {\path{doi:10.1021/acs.jctc.8b00832}}.

\bibitem{Zhang2018}
Y.~Zhang, D.~A. Kitchaev, J.~Yang, T.~Chen, S.~T. Dacek, R.~A. Sarmiento-Perez,
  M.~A.~L. Marques, H.~Peng, G.~Ceder, J.~P. Perdew, and J.~Sun.
\newblock \href{http://dx.doi.org/10.1038/s41524-018-0065-z}{Efficient
  first-principles prediction of solid stability: Towards chemical accuracy}.
\newblock {\em npj Comput. Mater.}, 4:9, 2018.
\newblock \href {https://doi.org/10.1038/s41524-018-0065-z}
  {\path{doi:10.1038/s41524-018-0065-z}}.

\bibitem{Nelder1965}
J.~A. Nelder and R.~Mead.
\newblock \href{http://dx.doi.org/10.1093/comjnl/7.4.308}{A {Simplex} {Method}
  for {Function} {Minimization}}.
\newblock {\em The Computer Journal}, 7:308--313, 1965.
\newblock \href {https://doi.org/10.1093/comjnl/7.4.308}
  {\path{doi:10.1093/comjnl/7.4.308}}.

\bibitem{Kacker2010}
R.~N. Kacker, R.~Kessel, and K.-D. Sommer.
\newblock \href{http://dx.doi.org/10.6028/jres.115.031}{Assessing differences
  between results determined according to the guide to the expression of
  uncertainty in measurement}.
\newblock {\em J. Res. Nat. Inst. Stand. Technol.}, 115(6):453, 2010.
\newblock \href {https://doi.org/10.6028/jres.115.031}
  {\path{doi:10.6028/jres.115.031}}.

\bibitem{Lejaeghere2014}
K.~Lejaeghere, J.~Jaeken, V.~V. Speybroeck, and S.~Cottenier.
\newblock \href{http://dx.doi.org/10.1103/physrevb.89.014304}{Ab initio based
  thermal property predictions at a low cost: An error analysis}.
\newblock {\em Phys. Rev. B}, 89:014304, jan 2014.
\newblock \href {https://doi.org/10.1103/physrevb.89.014304}
  {\path{doi:10.1103/physrevb.89.014304}}.

\bibitem{Lejaeghere2016}
K.~Lejaeghere, L.~Vanduyfhuys, T.~Verstraelen, V.~V. Speybroeck, and
  S.~Cottenier.
\newblock \href{http://dx.doi.org/10.1016/j.commatsci.2016.01.039}{Is the error
  on first-principles volume predictions absolute or relative?}
\newblock {\em Comput. Mater. Sci.}, 117:390--396, 2016.
\newblock \href {https://doi.org/10.1016/j.commatsci.2016.01.039}
  {\path{doi:10.1016/j.commatsci.2016.01.039}}.

\bibitem{Proppe2016}
J.~Proppe, T.~Husch, G.~N. Simm, and M.~Reiher.
\newblock \href{http://dx.doi.org/10.1039/c6fd00144k}{{U}ncertainty
  quantification for quantum chemical models of complex reaction networks}.
\newblock {\em Faraday Discuss.}, 195:497--520, 2016.
\newblock \href {https://doi.org/10.1039/c6fd00144k}
  {\path{doi:10.1039/c6fd00144k}}.

\bibitem{Proppe2017}
J.~Proppe and M.~Reiher.
\newblock \href{http://dx.doi.org/10.1021/acs.jctc.7b00235}{Reliable estimation
  of prediction uncertainty for physicochemical property models}.
\newblock {\em J. Chem. Theory Comput.}, 13:3297--3317, 2017.
\newblock \href {https://doi.org/10.1021/acs.jctc.7b00235}
  {\path{doi:10.1021/acs.jctc.7b00235}}.

\end{thebibliography}

\end{document}